\documentclass[aps,prb,twocolumn,superscriptaddress,showpacs]{revtex4-1}

\usepackage{graphicx}
\usepackage{amsmath}
\usepackage{color}
\usepackage{multirow}

\newcommand{\cbco}{CaBaCo$_4$O$_7$}

\begin{document}

\title{CaBaCo$_4$O$_7$: A ferrimagnetic pyroelectric}

\author{R. D. Johnson}\email{r.johnson1@physics.ox.ac.uk}
\affiliation{Clarendon Laboratory, Department of Physics, University of Oxford, Oxford, OX1 3PU, United Kingdom}
\author{K. Cao}
\affiliation{Department of Materials, University of Oxford, Parks Road, Oxford OX1 3PH, United Kingdom}
\author{P. G. Radaelli}
\affiliation{Clarendon Laboratory, Department of Physics, University of Oxford, Oxford, OX1 3PU, United Kingdom}

\date{\today}

\begin{abstract}
Magneto-electric coupling in pyroelectric {\cbco} is investigated using \emph{ab-intio} calculations and Landau theory. 
The former shows that exchange-striction is strong enough to produce a giant change in electric polarization upon ferrimagnetic ordering, comparable to the experimentally determined value of $\sim17~mC/m^{2}$. Furthermore, Landau theory demonstrates that magneto-elastic coupling in {\cbco} is responsible for the strong magneto-electric coupling appearing close to the magnetic phase transition.  
\end{abstract}


\maketitle

\section{\label{intro}Introduction}

Solid-state materials that adopt a polar crystal structure have sustained interest in condensed matter physics, and have become key components in technology. For example, the change in the intrinsic bulk electric polarization of non-centrosymmetric pyroelectric compounds, which occurs upon varying the temperature of the material, forms the basis of infra-red sensing devices. Also, research into ferroelectric materials, in which inversion symmetry is broken at a phase transition giving rise to switchable ferroelectric states, has lead to the development of electronic devices such as ferroelectric random access memory (FeRAM). Multiferroic materials form a subset of ferroelectrics, in which spontaneous electric polarization is coupled to long-range magnetic order. Research in this field has recently undergone a renaissance of interest, following the discovery of magnetic-field-switchable electric polarization in the now canonical systems TbMnO$_3$\cite{kimura03} and TbMn$_2$O$_5$ \cite{hur04} -- opening new routes towards the development of novel multifunctional devices.

To integrate multiferroic materials into technology it is necessary to identify systems that exhibit a very large magnetically-induced ferroelectric polarization close to room temperature.  However, the largest magnetically-induced ferroelectric polarization measured to date, (2870~$\mu$Cm$^{-2}$ observed in CaMn$_7$O$_{12}$ below 90~K  \cite{zhang11,johnson12,perks12}), is two orders of magnitude smaller than that of a good ferroelectric. Recently, a much larger spin-assisted change in polarization of $\sim17000~\mu$Cm$^{-2}$ was measured in  \cbco\ below 64~K\cite{singh12,caignaert13} --- a very significant observation that, if confirmed, could pave the way for a new generation of magnetic ferroelectrics.  In this paper, we perform first-principles calculations and a phenomenological analysis to study the magneto-electric coupling in \cbco. We demonstrate that all single crystal experimental data are consistent with  \cbco\  being \emph{pyro}electric rather than \emph{ferro}electric in both paramagnetic and magnetically ordered states.  The large pyroelectric currents observed near the magnetic phase transition result from an exchange-striction-driven change of ${\Delta P}$ in the paramagnetic pyroelectric polarisation ${P_{pyr}}$.  However,  ${\Delta P}$ is always co-aligned in a fix relation with ${P_{pyr}}$, either parallel or antiparallel depending on the sign of the magnetostrictive constant, and neither ${P_{pyr}}$ nor ${\Delta P}$ are switchable.

\section{\label{candm}Crystal and magnetic structures and electrical properties}

\begin{figure}
\includegraphics[width=8.8cm]{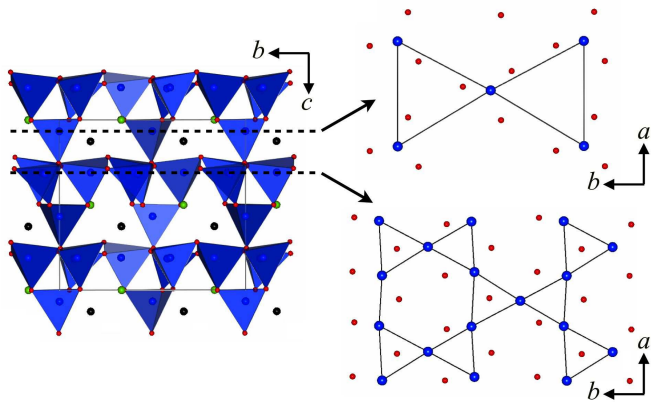}
\caption{\label{cryst}(Color online) Left: The crystal structure of \cbco. Calcium, barium, cobalt, and oxygen atoms are shown as black, green, blue, and red spheres, respectively. The CoO$_4$ tetrahedra are shaded blue. Right: The triangular and kagom$\mathrm{\acute{e}}$ CoO$_4$ layers in the $ab$ plane.}
\end{figure}

The crystal structure of \cbco, shown in Figure \ref{cryst}, was found to adopt the polar space group $Pbn2_1$ at all temperatures below 400~K \cite{caignaert09}. The structure comprises interleaved kagom$\mathrm{\acute{e}}$ and triangular layers of CoO$_4$ tetrahedra, which are buckled with respect to a high symmetry, high temperature trigonal polar phase (space group $P31c$) common to other members of the $R$BaCo$_4$O$_7$ series \cite{chapon06,huq06} ($R$ = rare earth, calcium or yttrium), but yet to be observed in \cbco. In both space group symmetries, \cbco\ is likely to be a non-switchable pyroelectric material, since atoms in inversion-related structures are separated by large distances. Furthermore, it follows that a high temperature phase transition to a centrosymmetric group is extremely unlikely to occur below the melting point.

\begin{figure}
\includegraphics[width=8.8cm]{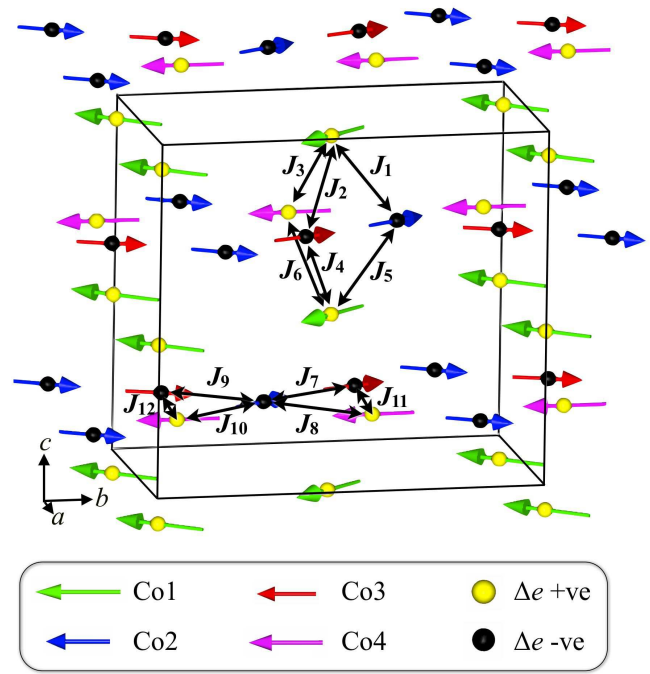}
\caption{\label{figmag}(Color online) The ground state magnetic structure of \cbco. Cobalt ions are shown as spheres colored yellow or black if charge rich or charge poor, respectively. Magnetic moments are colored in accordance with the scheme adopted in reference \citenum{caignaert10}. The twelve unique nearest neighbor exchange paths are shown by black arrows.}
\end{figure}

The geometric frustration intrinsic to both kagom$\mathrm{\acute{e}}$ and triangular lattices, well known to give rise to exotic frustrated magnetic ground states \cite{nakatsuji05,lee07}, is lifted as a result of the CoO$_4$ buckling. This structural distortion is reported to be largest in \cbco\ [\citenum{caignaert09}], removing the frustration, and promoting ferrimagnetic order developing at $T_\mathrm{c} = 64$ K. The magnetic structure\cite{caignaert09,caignaert10} is shown in Figure \ref{figmag}. There are four symmetry inequivalent cobalt sites in the unit cell, labelled Co1, Co2, Co3, and Co4, and colored green, blue, red and pink, respectively, in accordance with the color scheme in reference \citenum{caignaert10}. The magnetic moments of the four sites lie within the $ab$ plane with Co1 and Co4 moments approximately antiparallel to those of Co2 and Co3. \cbco\ is a mixed valance system, with greater charge, and hence larger magnetic moments, located on the Co1 and Co4 sites; a primary component to the ferrimagnetism. 

Measurements on a powder sample of \cbco\ [\citenum{singh12}] showed that a pyrocurrent signal, corresponding to a change in electrical polarization, coincided with an anomaly in the dielectric constant at the ferrimagnetic ordering transition $T_\mathrm{c}$. The pyroelectric signal switches sign with the external electric field, and the results were therefore interpreted as evidence for  ferroelectricity and multiferroicity\cite{singh12,caignaert13}. We note, however, that the magnetically-induced change in polarisation (${\Delta P} \sim 80 ~\mu$Cm$^{-2}$ was found to be extremely small (approximately 0.1 to 0.5\%) with respect to the ``pyroelectric polarization'' $P_{pyr}$, with whatever definition might be adopted for it (see Appendix I for an extended discussion);  these results therefore have to be interpreted with caution, as they could easily arise from an artefact.    Similar measurements on single crystal samples showed much larger pyroelectric currents  developing at $T_\mathrm{c}$, consistent with a giant change in polarisation of ${\Delta P} \sim17000~\mu$Cm$^{-2}$ \cite{caignaert13}. However, no switching behavior was reported for the single crystal sample.

\section{\label{meth}Methods}

Our first-principles calculations were based on density-functional theory implemented in the Vienna \emph{ab-initio} simulations package (VASP)\cite{kresse93,kresse96}. We used the spin-polarized generalized gradient approximation with on-site Coulomb interactions, $U$, included for cobalt 3d orbitals (GGA+U)\cite{liechtenstein95}. By fixing the Hund coupling constant $J=1$ eV and testing several $U$ values we found that the experimental ground state electronic structure becomes metallic if $U < 3$ eV. We therefore only present the results for $U=4$ eV, as \cbco\ is known to be an insulator. We also performed calculations with U=6 eV, which produced very similar results. The projector augmented-wave (PAW)\cite{blochl94} method with a 500 eV plane-wave cutoff was used throughout, and a $ 6 \times 4 \times 4$ $k$-point mesh converges the calculation very well. In calculations of relaxed crystal structures, first atomic positions, and then the lattice parameters, were left to vary until changes in total energy in the self-consistent calculations were less than 10$^{-7}$ eV and the remnant forces were less than 1 meV/\AA. The electric polarization was calculated using the Berry phase method\cite{king-smith93}.

\section{\label{pyro} Modelling Magnetism and Pyroelectricity}

We turn to the symmetric Heisenberg model to describe the magnetic interactions in \cbco. The expression for the magnetic Hamiltonian is simplified to $H_m = \sum_{ij}J_{ij}\boldsymbol{S}_i \cdot \boldsymbol{S}_j$,  where $J_{ij}$ are exchange integrals between cobalt spins $\boldsymbol{S}_i$ and $\boldsymbol{S}_j$.  For simplicity we only consider nearest neighbor (NN) interactions of which 12 are unique, labelled $J_1-J_{12}$ in Fig. \ref{figmag}. Furthermore, it is reasonable to assume that the main exchange-striction effects primarily originate in NN interactions, and are therefore proportional to $\boldsymbol{S}_i \cdot \boldsymbol{S}_j$.

The magnetic point group symmetry of \cbco\ below $T_\mathrm{c}$ ($m'm2'$) only allows a change in electric polarization parallel to the crystallographic $c$ axis to be induced by exchange-striction. All twelve exchange interactions can contribute to the change in electric polarization in the experimental ferrimagnetic (EFM) structure, which can be calculated directly. However, the paramagnetic (PM) phase presents a challenge, as spin polarized calculations are required to give accurate results. Thus, to mimic the PM phase, we averaged several spin configurations in such a way that individual NN exchange interactions cancel, $i.e.$, $\sum_{k} \boldsymbol{S}_i(k) \cdot \boldsymbol{S}_j(k)=0$, where $k$ denotes the spin configuration --- a method reliably used to calculate spin-phonon coupling in the PM phase of ZnCr$_2$O$_4$ [\citenum{fennie06}]. The average electric polarization between these spin configurations can then be considered an approximation of the polarization of the PM phase, as any magnetically-induced polarization will cancel, leaving only that intrinsic to the crystal structure. 

In \cbco\ the PM phase was modelled with four collinear spin configurations that we label FM, AFM1, AFM2, AFM3, respectively. FM is a ferromagnetic structure, while the other three are antiferromagnetic structures of the type (- - + +), (- + - +), (- + + -) for (Co1,Co2,Co3,Co4), respectively. The values of exchange interaction energies ($H_m$) for the four spin configurations are given in Table \ref{tab:exchange}, where each component exactly cancels within the Heisenberg spin model approximation.

\begin{table}
\begin{ruledtabular}
\caption{\label{tab:exchange}Atom pair, bond length ($l$), and exchange energies of a single unit cell for each spin configuration, given for all NN exchange interactions. Note that in the calculations, the magnetic moments and exchange parameters are normalized to 1.}
\begin{tabular}{ccccccc}
  J &  Co pair & $l$ (\AA) & FM & AFM1   &  AFM2   & AFM3    \\
\hline
$J_1$ & Co1-Co2 & 3.099 &4  & 4   &   -4   & -4     \\
$J_2$ & Co1-Co3 & 3.165 &4  & -4   &   4   & -4     \\
$J_3$ & Co1-Co4 & 3.107 &4  & -4   &   -4   & 4    \\
$J_4$ & Co1-Co3 & 3.187 &4  & -4   &   4   & -4        \\
$J_5$ & Co1-Co2 & 3.207 &4  & 4   &   -4   & -4   \\
$J_6$ & Co1-Co4 & 3.226 &4  & -4   &  -4   & 4    \\
$J_7$ & Co2-Co3 & 3.299 &4  & -4   &  -4   & 4   \\
$J_8$ & Co2-Co4 & 3.231 &4  & -4   &   4   & -4    \\
$J_9$ & Co2-Co3 & 3.007 &4  & -4   &   -4  & 4    \\
$J_{10}$ & Co2-Co4 & 3.162 &4  & -4 &  4    & -4     \\
$J_{11}$ & Co3-Co4 & 3.262 &4  & 4  &   -4  & -4  \\
$J_{12}$ & Co3-Co4 & 3.012 &4  & 4   &   -4   & -4   \\
\end{tabular}
\end{ruledtabular}
\end{table}

Having established a model for the PM phase, it is now possible to calculate and compare the electric polarization of the PM and EFM structures. Note that in the following the effect of spin-orbit coupling was found to be negligible, and so has been omitted from the final results. Initially we performed calculations on a fixed experimental atomic geometry\cite{caignaert10} in order to isolate the pure electronic contribution to the polarization. The results, given in the first part of Table \ref{tab:polar}, show only small variations for all spin configurations, with ${\Delta P}$ approximately 1 mC/m$^2$ -- one order of magnitude smaller than that measured by experiment\cite{caignaert13} -- indicating that the pure electronic contribution cannot fully explain the measured change in polarization observed upon magnetic ordering. Therefore, an ionic contribution to the electric polarization was calculated by fixing the experimental lattice parameters and relaxing the internal ion positions. The results are shown in the second part of Table \ref{tab:polar}, where it can be seen that both the relative fluctuations and the final ${\Delta P}$ are significantly enhanced. Finally, we allow for magnetic strain coupling by performing full relaxations under each magnetic configuration. The results, given in the third part of Table \ref{tab:polar}, show significant differences with respect to the electronic and ionic relaxation calculation, but give rise to a comparable ${\Delta P}$. Notably, the sign of the polarization in the EFM phase changes sign upon relaxation, however the relative change, ${\Delta P}$, remains positive.

The correct approximation of the PM phase is essential in evaluating ${\Delta P}$. To double check our model, we also generated a model PM state with a set of noncollinear spin configurations that satisfy the same cancellation relation. In this case, fully relaxing the atomic geometry gives ${\Delta P} = 4.6$ mC/m$^2$ with U=4 eV, which is in good agreement with that presented in Table \ref{tab:polar}. All calculations show that exchange-striction effects alone can give rise to a giant enhancement of the electric polarisation of the order 1-6 mC/m$^2$, comparable to the experimental observations. There remains a discrepancy between the actual magnitude of calculated and experimentally determined change in electric polarization. However, this effect is typical in such calculations due to the limitation of current DFT in the treatment of the strong correlation effects of, in this case, cobalt $3d$ electrons. Furthermore, the approximation of the paramagnetic phase required for calculations might reduce ${\Delta P}$ since the real PM phase is completely disordered, while our model is constructed from ordered states.

\begin{table}
\begin{ruledtabular}
\caption{\label{tab:polar}The calculated electric polarization in units mC/m$^2$ of the four spin configurations that average to give the net PM polarization, and that calculated for the experimentally determined ferrimagnetic structure (EFM). For each row, the polarization in the FM state is taken as a reference. ${\Delta P}= P(EFM)-P(PM)$.}
\begin{tabular}{cccccccc}
  FM &   AFM1   &  AFM2   & AFM3    & PM      & EFM    & ${\Delta P}$ \\
\hline
\multicolumn{8}{l}{\textit{Electronic only}} \\
 0   &   -0.16   & 0.36   & 1.06   &  0.31  & 0.74  & \bf{0.43} \\
\multicolumn{8}{l}{\textit{Electronic and ionic relaxation}} \\
 0   &   -14.0   & -10.4   & 1.1   &  -5.8  & -0.6   & \bf{5.2} \\
\multicolumn{8}{l}{\textit{Electronic, ionic, and lattice relaxation}} \\
0   &   -17.71   & -21.7   &  -3.1  &  -11.2  & -5.3   & \bf{5.9} \\
\end{tabular}
\end{ruledtabular}
\end{table}

\section{\label{pd}$\boldsymbol{H-T}$ phase diagram}

As explained in detail in Appendix I, the electrical polarisation is an ill-defined concept in a pyroelectric, because it depends upon the choice of a hypothetical centrosymmetric phase that is not actually present in the phase diagram.  However we chose to define it, the pyroelectric polarisation cannot be switched with an electric field, although one can reverse it by simply rotating the crystal upside down.  The key issue that remains to be addressed is whether ${\Delta P}$ is switchable, since a positive answer would indicate a very unusual kind of spin-assisted ferroelectricity.  In multiferroics, the magnetically-induced polarisation changes sign upon reversing the polarity of the magnetic structure, which must therefore be acentric.  For \cbco , reproducing this  mechanism is highly problematic:  although the crystal is highly polar, the cobalt sublattice is almost centrosymmetric and the cobalt magnetic structure is also quasi-centrosymmetric,  the inverted EFM structure being essentially identical to the EFM structure. For this very reason,  \emph{ab-intio} calculations alone are of limited use in assessing whether ${\Delta P}$ can switch, since there is no obvious way to construct a magnetic domain that would support a reversed ${\Delta P}$.  We therefore adopt a different approach based upon a phenomenological Landau theory, and demonstrate that ${\Delta P}$ does not switch.   Furthermore, this minimal model is found to be sufficient to describe the experimental  behavior of \cbco\ when tuned by temperature and magnetic field close to $T_\mathrm{c}$.  We will employ a scalar notation throughout.  Extension to tensor notation is straightforwards but does not add to the essential physics of the problem.

We start by defining a simple Landau free energy for the paramagnetic phase that accounts for its electrical properties in the vicinity of the equilibrium pyroelectric structure.  We can write the free energy as:

\begin{equation}
\label{eq: param_landau}
F_p=-\frac{\alpha}{2} P^2+\frac{\beta}{4}P^4-PE
\end{equation}

where $\alpha$ and $\beta$ are positive constants, which entirely determine the polarisability $\chi_0$ an hyper-polarisability $h_0$ of the paramagnetic phase --- both measurable quantities:

\begin{eqnarray}
\label{eq: param_diahyp}
\chi_0&=&\left . {\frac{\partial P}{\partial E}} \right \vert_{E=0}=\frac{1}{2 \alpha} \nonumber\\
h_0&=&\frac{1}{2}\left . {\frac{\partial^2 P}{\partial E^2}}\right \vert_{E=0}=\mp \sqrt{ \frac{9}{2}\chi_0^5 \beta}
\end{eqnarray}

Since we are only interested in the behaviour close to the magnetic transition, we ignore the temperature dependences of $\alpha$ and $\beta$, which would give rise to conventional pyroelectric currents upon heating.  From eq. \ref{eq:  param_landau} we can also extract the equilibrium value of $P$:

\begin{equation}
\label{eq:  P0_defin}
P_0=\pm \sqrt{\frac{\alpha}{\beta}}
\end{equation}

We stress that $P_0$ is \emph{not} a measurable quantity and does not correspond to a density of dipole moments $P_{pyr}$ estimated using any of the reference structures proposed in Appendix I.  In fact, this would entail extending eq.  \ref{eq:  param_landau} much beyond the range in which the quartic approximation is valid.  However, changes in $P$ in the vicinity of the equilibrium position are well defined and would result in a measurable current in the standard experimental setup.  Below we present an estimation of $P_0$ from \emph{ab initio} calculations. 

We first consider how the Landau free energy in eq. \ref{eq:  param_landau} can be modified in the magnetically ordered phase.  The point group symmetry of \cbco\ changes from $mm2.1'$ to $m'm2'$ at the magnetic ordering transition. Only time reversal symmetry is broken, with all spatial symmetry operations preserved. The lowest order magneto-electric coupling invariant in the Landau expansion of the free energy is $\gamma P M^2$, where $\gamma$ is a coupling constant, $P=P_0 + \Delta P$, and $\Delta P$ is the change in electric polarization upon magnetic ordering as before.  $M$ is the magnetic order parameter, which, although primarily antiferromagnetic, is coupled with the magnetic field through the ferrimagnetic component. $\gamma P M^2$ is time reversal even and parity odd, and as such is allowed \emph{prima facie} given the polar, paramagnetic parent phase point group. However, the Landau free energy must also be invariant by any continuous or discrete free-space operator applied to the crystal as a whole.  If one applies an inversion operator (not part of the crystal symmetry) the magnetic structure remains almost invariant, as explained above, while $P$ changes sign.  Since this is an approximate relation, we can only conclude that the $\gamma P M^2$ term must be very small.  An alternative, and perhaps more intuitive interpretation is to consider that \cbco\ undergoes a hypothetical non-polar to polar structural phase transition to one of the reference structures at very high temperatures.  The term $\gamma P M^2$ is not invariant in any of the hypothetical parent phases, and is therefore rigorously excluded.

We proceed to demonstrate that a change in bulk electric polarization can occur via a magneto-elastic contribution to the free energy described by the higher order term $-\frac{c}{2} M^2 P^2$, which can result in a fractional change ${\Delta P}$ of the electric polarization.

The lowest order, stable Landau expansion of the free energy may be written as:
\begin{eqnarray}
\label{fe}
&F(M,P,B) =  F_0 - BM + \frac{a}{2}(T-T^*)M^2 +\frac{b}{4}M^4 \nonumber\\
& -\frac{c}{2}M^2P^2 + \frac{d}{6}M^6 -\frac{\alpha}{2} P^2+\frac{\beta}{4}P^4-PE
\end{eqnarray}

where $a$, $b$ and $d$ and $T^*$ are constants of the purely magnetic part of the free energy. $c$ is the magneto-elastic coupling constant that may be positive or negative, and all other constants are defined such that positive values stabilize the free energy.

In zero applied electric field, two equilibrium conditions follow:
\begin{equation}
\label{eosM}
\frac{\partial F}{\partial{M}} = -B +a(T-T^*)M + b{M}^3 -c{P}^2{M} + d{M}^5 = 0
\end{equation}
and,
\begin{equation}
\label{eosP}
\frac{\partial F}{\partial{P}} = -\alpha{P} + \beta{P}^3 -c{M}^2{P} = 0
\end{equation}

Substiting eq. \ref{eq:  P0_defin} into equation \ref{eosP} and rearranging the terms gives the solution

\begin{equation}
\label{p}
{P}^2 = {P_0}^2 +\frac{c}{\beta}{M}^2
\end{equation} 
where ${M}$ is found by solving equation \ref{eosM}, as described below.  Equation \ref{p} contains the essence of the physics of \cbco:  at the magnetic ordering temperature, ${M}$ becomes non-zero, and an additional contribution to the polarisation ${\Delta P}$ develops due to magneto-striction. The sign of ${\Delta P}$ in relation to ${P_0}$ is fixed once and for all by the sign of the coupling constant $c$ and can never be switched.

In principle, it is possible to determine the values of $\alpha$ and $\beta$, and therefore derive $P_0$ from \emph{ab initio} calculations of the paramagnetic phase in applied electric field.  However, the aforementioned difficulty of performing accurate calculations in the paramagnetic phase makes this approach impractical.  We therefore chose a different method, exploiting the fact that $P_0$ enters as a parameter in the free energy expansion as a function of $\Delta P$ around the magnetic ground state.  As before, all calculations here are performed with a Coulomb interaction energy of U=4 eV.  We first estimate the \emph{ab initio} atomic structure of the paramagnetic phase (corresponding to $\Delta P=0$) by averaging  the relaxed atomic structures in the four spin configurations (FM, AFM1, AFM2, AFM3) with fixed experimental lattice parameters.   The ground-state atomic structure (corresponding to $\Delta P=\Delta P_{max}$) was previously determined by relaxing the atomic positions with the ground-state magnetic structure (EFM).  By interpolating between these two extremes we can estimate atomic positions, and therefore calculate ground-state energies, as a function of $\Delta P$.  In these calculations, it was assumed that $M$ does not vary, a reasonable approximation for values of ${\Delta P}$ close to the ground state. These energies were then fitted to the expression $E=E_0+ a^\prime ({ P_0}+{\Delta P})^2+b^\prime ({P_0}+{ \Delta P})^4$, which closely follows the form of our Landau theory.

\begin{figure}
\includegraphics[width=8cm]{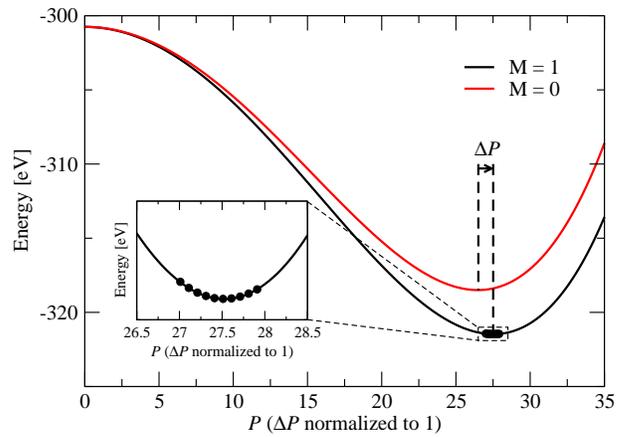}
\caption{\label{ecalcfig}Fit of the polarization free energy to ${P}$, determined at points close to the energetic minimum by \emph{ab-initio} calculations. The energy for ${M} = 0$ and ${M} = 1$ is shown by red and black lines, respectively.}
\end{figure}

The results of the calculations and are plotted as closed circles in Figure \ref{ecalcfig} as a function of $P/ \Delta P_{max}$, together with the fitted expression, plotted as a line.  From the fit, we extract a value $P_0/\Delta P_{max}=26.51$. Taking the reported value of $\Delta P_{max}=17$ mC/m$^{2}$, this gives $P_0 \simeq 450$ mC/m$^{2}$  We note that the calculated $E$v$P$ curve is symmetrical about the origin (not shown in Figure \ref{ecalcfig}). Hence, there exists a single energy minima for each $\pm {P_0}$ domain, respectively. The energy needed to reverse ${\Delta P}$, i.e. switch between energetic minima, while preserving the magnetization is about 411 meV, consistent with a value of 464 meV from direct \emph{ab-initio} calculations This result clearly indicates that ${\Delta P}$ is not switchable for a given ${P_0}$ domain.

\begin{figure}
\includegraphics[width=8.5cm]{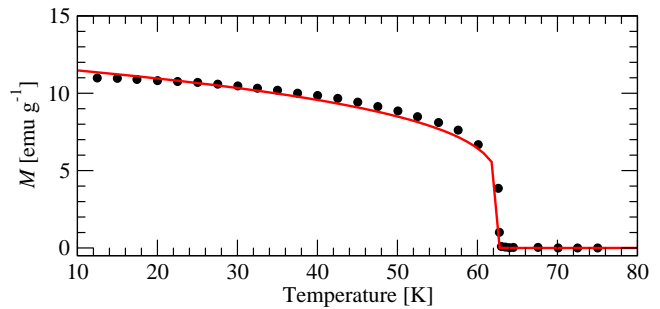}
\caption{\label{tdeps}(Color online) The temperature dependence of the magnetization in zero applied magnetic field. Data points (black circles) have been extracted from reference \citenum{caignaert13}. The fit to the data, described in detail in the text, is shown as a red line.}
\end{figure}

We now turn to the magnetization. Substituting equation \ref{p} into equation \ref{eosM} gives
\begin{equation}
\label{eqmag}
-{B} +a(T-T_c){M} + \tilde{b}{M}^3+ d{M}^5 = 0
\end{equation}
where $T_c = T^* + \frac{c}{a}{P_0}^2$ and $\tilde{b} = b - \frac{c^2}{\beta}$. These expressions capture two further key aspects of the physics of \cbco. Coupling to the pyroelectric polarization present in the paramagnetic phase will increase the magnetic ordering temperature, and furthermore, may induce a first-order (negative $\tilde{b}$) magnetic phase transition, as opposed to a second-order transition (positive $\tilde{b}$) expected in magnetic systems. Magnetization data measured parallel to the $b$ axis for ${B}\simeq 0$ has been reproduced from reference \citenum{caignaert13} (Figure \ref{tdeps}), which in the following we take to represent the thermal evolution of the magnetic order parameter. Despite no evidence for magnetic hysteresis at the phase transition \cite{singh12}, there occurs a sharp jump in the magnetization at $T_c$ --- evidence for first-order behavior. This was confirmed by fitting equation \ref{eqmag} to the data, in units of emu/g, and with ${B}=0$. The constants $\tilde{b}$ and $d$ were allowed to vary freely in the fit, with $a$ set to unity having factored out a scaling parameter, $\xi$, to be determined later. The best fit is shown in Figure \ref{tdeps}, with $a = \xi$, $\tilde{b}  =  -0.112\xi$, $d = 0.00385\xi$, and $T_c = 62.0$~K. We note that $\tilde{b}$ takes a negative value, indicating that the magnetic phase transition is indeed first order due to coupling to $P$. This result might be verified experimentally by investigating magnetic hysteresis. 

\begin{figure}
\includegraphics[width=8.5cm]{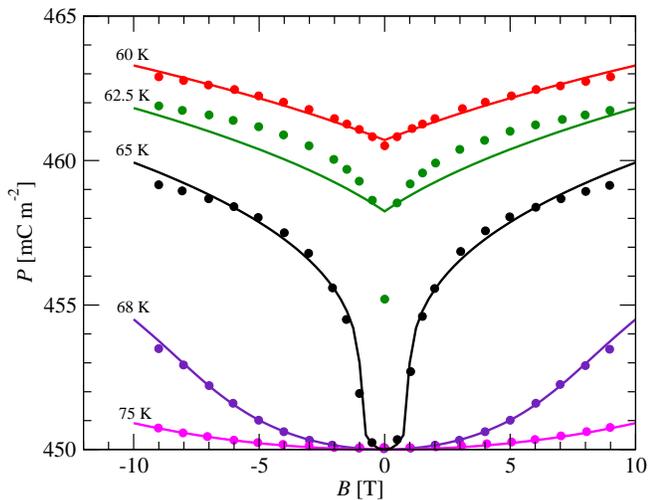}
\caption{\label{bdep}(Color online) The magnetic field dependence on electric polarization at four temperatures close to the phase transition. Data points (circles) have been extracted from reference \citenum{caignaert13}. Simulations of the data are shown as solid lines, which are colored according to the respective temperatures. Note that $P_0$ has been added throughout.}
\end{figure}

Having established the temperature dependence of ${M}$, setting the scaling parameter $\xi$ to 0.4 and the ratio of constants $c$ and $\beta$ that couple ${M}$ and $P$ to $\frac{c}{\beta} = 210$, gave the best qualitative agreement with the magnetic field dependence of the electric polarization (Figure \ref{bdep}). This result clearly demonstrates that the experimentally determined $H$-$T$ phase diagram close to $T_\mathrm{c}$ can be explained by our phenomenological model based solely on magneto-elastic coupling. 

Finally, we again consider the energy barrier associated with the hypothetical switching of ${\Delta P}$. The difference in energy of two $\pm\Delta{P}$ ferroelectric domains can be written as:
\begin{eqnarray}
\Delta E&=&-\frac{c}{2}{M}^2({P_0}+\Delta {P})^2+\frac{c}{2}{M}^2({P_0}-\Delta{P})^2\nonumber\\
&=&-2c {P_0}P{M}^2
\end{eqnarray}
\emph{i.e.} there is a large energy barrier to switching $\Delta{P}$ that scales with ${P_0}$, consistent with the results of the calculations described in the above.

\section{\label{conc}Conclusions}
To summarize, we have demonstrated through \emph{ab-initio} calculations that in \cbco, the giant change in electric polarization observed at the phase transition from pyroelectric paramagnetic to pyroelectric ferrimagnetic can arise as a result of exchange-striction effects alone. The change in polarization was found to be an enhancement, $i.e.$ in the same direction as the polarization of the paramagnetic phase, and not switchable. Furthermore, such large changes are only predicted when one considers the relaxation of ionic positions. Our \emph{ab-initio} results are supported by Landau theory, which predicts the correct magneto-electric behavior apparent close to $T_\mathrm{c}$.

\section{\label{p0}Appendix I:  Absolute value of electric polarization}

Experimentally, values of electric polarization may be measured in two ways. Firstly, relative changes in polarization can be determined by integrating a pyroelectric current measured upon tuning the system in question between two different structures along a given \emph{path}, discussed below. Secondly, the absolute polarization may be measured if the system is ferroelectric, and domain switching has been observed. Modern theories of electric polarization, where values can be quantum mechanically calculated using the Berry phase method \cite{king-smith93}, closely follow the pyroelectric experimental procedure. The theory is therefore limited to calculating relative changes in electric polarization as follows. The Berry phase method shows that the electric polarization is multivalued with a period of $\frac{1}{\Omega}e\boldsymbol{R}$, where $\Omega$ is the unit cell volume, $e$ is the electron charge, and $\boldsymbol{R}$ is any lattice vector. Therefore, to define an absolute spontaneous polarization in a non-centrosymmetric material, it is necessary to determine the origin of the above periodic polarization. This may be constrained by designating a `nearest' (smallest total displacement of atoms) high temperature centrosymmetric structure with zero net polarization, e.g. the cubic phase of BaTiO$_3$. In order to calculate an absolute value of ${P_{pyr}}$ in \cbco, it was therefore necessary to define a hypothetical, high temperature crystal structure with zero net polarization - a centrosymmetric supergroup of $Pbn2_1$. To find such structures, we used the PSEUDO code of the Bilbao Crystallographic Server \cite{capillas11}, which allows one to determine the nearest supergroup structure for an arbitrary subgroup structure. With the experimental atomic structure as input \cite{caignaert10}, two suitable supergroup structures were identified with space groups $Pnna$ and $Pccn$, and the structural parameters are given in Table \ref{tab:refstruc}. In terms of atomic displacements, the $Pnna$ structure is the nearest.

\begin{table}
\begin{ruledtabular}
\caption{\label{tab:refstruc}Atomic positions of the $Pnna$ and $Pccn$ reference structures, given in fractional coordinates of the subgroup setting. $u_x$,$u_y$,$u_z$ are the displacement vectors that generate the experimental $Pbn2_1$ structure from the centrosymmetric structures with the origin shifted by (0, 0, -0.2) and (0, 0, -0.44) for $Pnna$ and $Pccn$, respectively.}
\begin{tabular}{ccccc}
 $\boldsymbol{Pnna}$\\
  Atom &  Fractional Coordinates   & $u_x$  & $u_y$   & $u_z$      \\
         Ca &    (0.0000,    0.67290,    0.7500)  &        0.0066    &   0.00000    &  -0.0792        \\
         Ba &    (0.0000,    0.66510,    0.2500)  &        0.0043    &   0.00000    &   0.0500       \\
         Co1 &   (0.0000,    0.99970,    0.7500)  &        0.0155   &    0.00000   &   -0.0168        \\
         Co2 &   (0.8703,    0.13215,    0.4999)  &        0.1230   &    0.03665   &   -0.0097        \\
         Co3 &   (0.1297,    0.13215,    0.0001)  &        0.1230   &   -0.03665  &    -0.0097        \\
         Co4 &   (0.2673,    0.00000,    0.5000)  &        0.0000   &   -0.07910  &    -0.0159        \\
         O1  &   (0.9845,    0.00000,    0.0000)  &        0.0000   &    0.00440   &    0.0460       \\
         O2  &   (0.9955,    0.50000,    0.0000)  &        0.0000   &   -0.00860  &     0.0285       \\
         O3  &   (0.7500,    0.25000,    0.5000)  &        0.0317   &    0.01130   &    0.0815       \\
         O4  &   (0.7500,    0.75000,    0.0000)  &       -0.0181  &    -0.00820  &     0.0141       \\
         O5  &   (0.0000,    0.15220,    0.2500)  &       -0.0549  &     0.00000   &    0.0497       \\
         O6  &   (0.2340,    0.08130,    0.7530)  &       -0.0305  &     0.02840   &    0.0532       \\
         O7  &   (0.2340,    0.91870,    0.2470)  &        0.0305   &    0.02840   &    0.0532       \\
  \\
   $\boldsymbol{Pccn}$\\       
         Atom &  Fractional Coordinates   & $u_x$  & $u_y$   & $u_z$     \\
         Ca &    (0.0000, 0.75000, 0.50000)  &    0.0066    &   -0.07710   &   -0.06920       \\
         Ba &    (0.0000, 0.75000, 0.00000)  &    0.0043    &   -0.08490   &    0.06000      \\
         Co1 &   (0.1241, 0.96030, 0.37455)  &   -0.1086   &     0.03940    &   0.11865       \\
         Co2 &   (0.8703, 0.28665, 0.24990)  &    0.1230    &   -0.11785   &    0.00030       \\
         Co3 &   (0.1297, 0.21335, 0.75010)  &    0.1230    &   -0.11785   &    0.00030       \\
         Co4 &   (0.3759, 0.96030, 0.12545)  &   -0.1086   &    -0.03940   &    0.11865       \\
         O1  &   (0.9900, 0.99790, 0.75875)  &   -0.0055   &     0.00650    &   0.04725       \\
         O2  &   (0.9900, 0.49790, 0.74125)  &    0.0055    &   -0.00650   &    0.04725       \\
         O3  &   (0.7891, 0.32580, 0.38765)  &   -0.0074   &    -0.06450   &   -0.04615        \\
         O4  &   (0.7500, 0.74180, 0.75000)  &   -0.0181   &     0.00000    &   0.02410       \\
         O5  &   (0.8548, 0.10255, 0.99975)  &    0.0903    &    0.04965    &   0.05995       \\
         O6  &   (0.2109, 0.17420, 0.61235)  &   -0.0074   &    -0.06450   &   -0.04615        \\
         O7  &   (0.3548, 0.89745, 0.00025)  &   -0.0903   &     0.04965    &   0.05995       \\
\end{tabular}
\end{ruledtabular}
\end{table}

Unfortunately, calculations showed that both the two supergroup structures are metallic, preventing \emph{ab-initio} Berry phase calculations. To obtain a qualitative estimation of ${P_{pyr}}$, we therefore assume the two reference structures are insulating and use point charge models to calculate the electric polarization, i.e. $P=\sum_iZ_i {\bf r}_i/ \Omega$, where ${\bf r}_i$ is the position of the $i$th ion, and $Z_i$ is the corresponding effective charge. Using simple valence charges with $Z$(Ca)=$Z$(Ba)=+2, $Z$(Co)=+2.5, $Z$(O)=-2, the absolute value of polarization of the PM phase is estimated to be  -775 mC/m$^2$ for $Pnna$ and  263 mC/m$^2$ for $Pccn$.
 
At first sight, it is surprising that two reference structures give completely different values for ${P_{pyr}}$. However, the displacement patterns shown in Table \ref{tab:refstruc} provide an explanation for this discrepancy. In the following we consider only the displacement along the $c$ direction, as the $ab$ plane components cancel exactly by symmetry. For $Pnna$, almost all cations move in an opposite direction to the anions, giving rise to the large polarization difference along the $+c$ direction. By comparison, in the case of $Pccn$, cations and anions are both displaced along the positive and negative $c$ directions, resulting in polar contributions that partially cancel with each other, giving rise to the relatively small polarization difference with a sign opposite to that for $Pnna$.

\begin{figure}
\includegraphics[width=8.5cm]{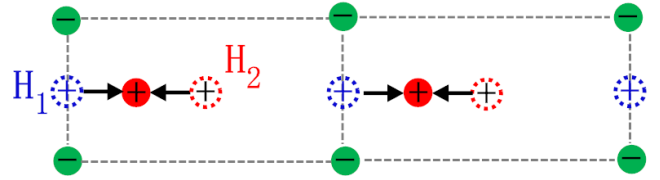}
\caption{\label{probfig}(Color online) 2D illustration of the ambiguity associated with choosing the correct path from centrosymmetric reference structure to the polar phase. Here we fix the position of the anions and use cations to represent different phases, where solid red represents the polar phase, dotted blue for the H$_{1}$ reference and dotted red for the H$_{2}$ reference. Black arrows denote directions of polar distortion.}
\end{figure}

A simple analogy can be drawn if one considers a 2D lattice of positive and negative charges, illustrated in Figure \ref{probfig}. Here, it is clear how two opposite displacements of charge can result in different `nearest' centrosymmetric structures, giving rise to an ambiguity associated with the choice of non-polar reference structures. Although H$_2$ appears energetically more favourable than H$_{1}$ in the 2D example, the case of \cbco\ is much more complex making it difficult to determine the correct reference structure. In general, real ferroelectric distortions are much smaller than those considered in this paper, so it is often trivial to identify the nearest reference structure, and the polarization may be determined absolutely if the parent phase crystal structure is known. Despite both hypothetical centrosymmetric structures of \cbco\ having unphysical structural distortions, careful variable-temperature experiments could be performed to identify the direction of the pyroelectric polarization with respect to the crystal structure, and hence remove the ambiguity in determining ${P_{pyr}}$.

\begin{acknowledgments}
This work was funded by an EPSRC grant, number EP/J003557/1, entitled ``New Concepts in Multiferroics and Magnetoelectrics".
\end{acknowledgments}

\bibliography{cbco,multiferro}

\end{document}